\documentclass[%
reprint,
superscriptaddress,
 amsmath,
 amssymb,
 aps,
 amsfonts,
prl,
]{revtex4-2}

\usepackage{graphicx}
\usepackage{dcolumn}
\usepackage{mathtools,slashed}
\usepackage[T1]{fontenc}
\usepackage{bm}
\usepackage{physics}
\usepackage{soul}
\usepackage{array}
\usepackage{pifont}         
\usepackage{booktabs}       
\usepackage{hyperref}
\hypersetup{
    colorlinks=true,
    linkcolor=blue,
    filecolor=magenta,      
    urlcolor=blue,
    citecolor=blue,
    pdftitle={Overleaf Example},
    pdfpagemode=FullScreen,
    }

\begin{document}
\title{Geometric Origin of Phonon Magnetic Moment in Dirac Materials}
\author{Wenqin Chen}
\affiliation{Department of Physics, University of Washington, Seattle, Washington 98195, USA}
\affiliation{Theoretical Division, T-4 and CNLS, Los Alamos National Laboratory, Los Alamos, New Mexico 87545, USA}
\author{Xiao-Wei Zhang}
\affiliation{Department of Materials Science and Engineering, University of Washington, Seattle, Washington 98195, USA}
\author{Ting Cao}
\affiliation{Department of Materials Science and Engineering, University of Washington, Seattle, Washington 98195, USA}
\author{Shi-Zeng Lin}
\email{szl@lanl.gov}
\affiliation{Theoretical Division, T-4 and CNLS, Los Alamos National Laboratory, Los Alamos, New Mexico 87545, USA}
\affiliation{Center for Integrated Nanotechnologies (CINT), Los Alamos National Laboratory, Los Alamos, New Mexico 87545, USA}
\author{Di Xiao}
\email{dixiao@uw.edu}
\affiliation{Department of Materials Science and Engineering, University of Washington, Seattle, Washington 98195, USA}
\affiliation{Department of Physics, University of Washington, Seattle, Washington 98195, USA}

\begin{abstract}
    We develop a theory for the phonon magnetic moment in doped Dirac materials, treating phonons as emergent gauge and gravitational fields coupled to Dirac fermions in curved space. By classifying electron-phonon coupling into angular momentum channels of Fermi surface deformation, we show that the phonon moment arises from two mechanisms: proportional to the electron Hall conductivity through the emergent gauge field coupling, and to the Hall viscosity through the frame field coupling. Applying our theory to Cd$_3$As$_2$ with first-principles calculations, we find quantitative agreement with experiment. Our results reveal a general mechanism for dynamically generating large phonon magnetism in metals and suggest a new route for probing Hall viscosity via phonon dynamics.
\end{abstract}
\maketitle

\textit{Introduction.}---Phonons, the quantized vibrations of a crystal lattice, are among the most fundamental bosonic excitations in solids. Traditionally regarded as non-magnetic, phonons have recently been shown to carry orbital magnetic moments, opening up an exciting avenue for exploring novel transport and optical phenomena~\cite{Chiral-Phonons-at-High-Symmetry,Observation-of-chiral,Truly-chiral,Chiral-phonons-in-quartz,Giant-thermal,Phonon-Thermal,Chiral-phonons-in-the,Phonon-Angular-Momentum-Hall,Thermal-Hall-conductivity,Large-phonon-thermal-Hall,Berry-Phase-of-Phonons,Phonon-Hall-Viscosity,Frequency-Splitting-of-Chiral-Phonons,Adiabatic-Dynamics-of-Coupled-Spins,Conversion-between-electron-spin,An-effective-magnetic-field,Large-effective-magnetic,Terahertz-electric-field,Phononic-switching-of-magnetization,Phono-magnetic-analogs,Giant-effective-magnetic-fields,Phonon-driven-spin-Floquet,Light-Driven-Spontaneous,Dynamically-induced-magnetism}.  Early theoretical work based on a classical picture of circulating ions predicted small phonon moments on the order of the nuclear magneton due to the heavy mass of ions~\cite{Dynamical-multiferroicity,Orbital-magnetic}. However, recent experiments have reported much larger moments---approaching the Bohr magneton---in a wide range of materials~\cite{A-Large-Effective,Magnetic-Control,Observation-of-interplay,Fluctuation-enhanced-phonon-magnetic-moments,Exciton-activated-effective-phonon-magnetic-moment,Origin-of-Large-Effective-Phonon-Magnetic}. These findings have motivated the development of quantum theories that attribute the magnetic moment to electron-phonon coupling~\cite{Geometrodynamics-of-electrons,Geometric-orbital-magnetization,Phonon-Magnetic-Moment,Gate-Tunable-Phonon,Giant-effective-magnetic-moments,Quantum-Nonlinear-Acoustic-Hall}, though these theories have thus far focused primarily on insulating systems.

Remarkably, large phonon moments have also been observed in metallic systems such as doped Cd$_3$As$_2$~\cite{A-Large-Effective} and Pb$_{1\text{-}x}$Sn$_x$Te~\cite{Observation-of-interplay}. Cd$_3$As$_2$ is a Dirac semimetal with bulk Dirac fermions, while Pb$_{1-x}$Sn$_x$Te is a topological crystalline insulator featuring gapless Dirac surface states. These observations suggest a strong connection between Dirac fermions and the emergence of phonon magnetic moments in gapless systems, yet the underlying mechanism remains poorly understood.

In addition to their experimental relevance, Dirac fermions allow an elegant framework for modeling electron-phonon coupling. Previous studies have shown that acoustic phonons and lattice strain can couple to electrons as emergent gauge and gravitational fields~\cite{Geometrodynamics-of-electrons,Phonons-and-electron-phonon,Gauge-fields-in-graphene,Elastic-Gauge-Fields-in-Weyl,Chiral-Anomaly-from-Strain-Induced,Inhomogeneous-Weyl-and-Dirac-Semimetals,Quantum-Nonlinear-Acoustic-Hall,Torsional-Response-and-Dissipationless-Viscosity,Dissipationless-phonon-Hall,Viscoelastic-response-of-topological-tight-binding,Circular-Phonon-Dichroism}. More recently, it has been recognized that optical
phonons can also couple to Dirac fermions as emergent
gauge fields~\cite{Symmetry-based-approach-to-electron-phonon,Hall-viscosity-for-optical-phonons,Phonon-Helicity}, giving rise to a phonon magnetic
moment~\cite{Gauge-theory-of-giant-phonon}.

In this work, we demonstrate that optical phonons can
also act as gravitational fields. Building on the frame-
work of Dirac fermions in curved space, we develop a
theory for the phonon magnetic moment in doped Dirac
materials. By classifying electron-phonon coupling into angular momentum channels of Fermi surface deformation, we link the orbital magnetic moment of phonons to electronic transport coefficients. A key result is that optical phonons can act as a dynamic frame field, inducing a phonon magnetic moment through feedback from the Hall viscosity of the electron fluid. Combined
with first-principles calculations, we apply our theory to Cd$_3$As$_2$, which yields a phonon magnetic moment consistent with experiment~\cite{A-Large-Effective}.  Our results not only provide a microscopic explanation for giant phonon moments in metals, but also suggest a route for experimentally probing Hall viscosity via phonon dynamics.

\begin{figure}
    \centering
    \includegraphics[width=\linewidth]{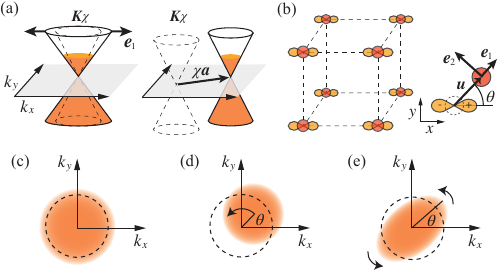}
    \caption{Schematic illustration of phonon-induced gauge and frame fields in Dirac materials. (a) Emergent gauge field $\chi \vb{a}$ shifts the Dirac point, while frame field $e_\mu^A$ distorts the Dirac cone. (b) In an $sp$-orbital lattice model, the frame field $(\vb{e}_1, \vb{e}_2)$ arises from bond stretching due to phonon displacement $\vb{u}$. Angular momentum channels contributing to the phonon magnetic moment: (c) Trivial monopolar ($l=0$) channel; (d) Dipolar ($l=1$) channel: phonon displaces the Fermi surface and drives circular motion; (e) Quadrupolar ($l=2$) channel: phonon distorts the Fermi surface into a nematic shape and rotates it.}
    \label{fig1}
\end{figure}

\textit{Geometrization of phonons.}---We begin by developing a geometric framework in which phonons act as frame fields and gauge fields in the low-energy theory of Dirac materials, effectively “geometrizing” lattice vibrations. For simplicity, we consider spinless Dirac points in three dimensions, described by the low-energy Hamiltonian $\mathcal{H}_D = v_F \vb{k} \cdot \boldsymbol{\sigma}$, where $\boldsymbol{\sigma}$ are the Pauli matrices, $v_F$ is the Fermi velocity, and $\vb{k}$ is the electron momentum. The effect of spin will be addressed later when we apply our theory to Cd$_3$As$_2$. Due to the linear band crossing, small perturbations cannot open a gap at the Dirac point; instead, they can only (i) shift its position in momentum space and (ii) modify the anisotropy or magnitude of the Dirac dispersion, as illustrated in Fig.~\ref{fig1}(a). These effects are naturally encoded in a geometric language: the momentum shift corresponds to a minimal coupling to an emergent gauge field $\vb{a}$, while the velocity renormalization corresponds to a frame field (vielbein) $e^{\mu}_A$ that acts as an emergent gravitational field by rescaling the local dispersion.  Including these phonon-induced fields along with the external electromagnetic vector potential $\vb{A}$, the system is described by the Dirac Hamiltonian in curved space~\cite{Weyl1929,Fock1929,Gauge-fields-in-graphene},
\begin{equation}
\label{effective hamiltonian}
\mathcal{H}_{\text{eff}} = v_F \sum_{\mu, A} (k_{\mu} - \chi a_{\mu} - A_{\mu}) e^{\mu}_A \sigma^A - \varepsilon_F,
\end{equation}
where $\varepsilon_F$ is the Fermi energy, $\chi = \pm 1$ denotes the valley index, $\mu = x, y, z$ labels global Cartesian coordinates, and $A = 1, 2, 3$ indexes the local tangent space spanned by the frame field.  In what follows, we demonstrate that both gauge and frame fields naturally arise from optical phonons. For concreteness, we focus on zone-center phonon modes ($\vb{q} = 0$) confined to the $xy$ plane.

To build intuition for the less-familiar concept of a frame field, we begin with a heuristic argument illustrating how it can be parameterized in terms of phonon modes. In a lattice distorted by a phonon mode, we can define a set of local frame vectors $(\vb{e}_1, \vb{e}_2)$ based on the atomic displacement field $\vb{u}$. Specifically, $\hat{e}_1$ is chosen to align with the direction of $\vb{u}$, while $\hat{e}_2$ is orthogonal to it.  The transverse vector $\vb{e}_2 = \hat{e}_2$ remains a unit vector, but the longitudinal vector $\vb{e}_1$ is stretched or compressed by the displacement, becoming $\vb{e}_1 = (1 - \beta u/a)\hat{e}_1$, where $a$ is the lattice constant and $\beta$ is a dimensionless parameter characterizing the strength of the distortion. Decomposing the local frame vectors into the global Cartesian basis yields the components of the frame field tensor $e^\mu_A$:
\begin{equation}
\label{frame field}
\begin{aligned}
e^{x}_1 &= \qty(1-\frac{\beta u}{a})\cos\theta,\quad e^{x}_2 = -\sin\theta,\\
e^{y}_1 &= \qty(1-\frac{\beta u}{a})\sin\theta,\quad e^{y}_2 = \cos\theta,
\end{aligned}
\end{equation}
where $\theta$ is the instantaneous angle between the displacement vector $\vb{u}$ and the $x$-axis. In this way, lattice vibrations dynamically modify the local geometry experienced by electrons, with the phonon displacement field entering the low-energy theory through the frame field.

With this intuitive understanding in hand, we now turn to a microscopic analysis of how optical phonons induce both emergent frame and gauge fields. To that end, we examine a tight-binding model defined on a cubic lattice [Fig.~\ref{fig1}(b)], with two orbitals per site. The Hamiltonian takes the form
\begin{equation}
    \label{tight-binding}
    \begin{aligned}
        H &= \frac{1}{2}\sum_{j}\sum_{\mu=x,y} \left[\psi^{\dagger}_{j+\vb{a}_{\mu}}(it\sigma^{\mu} - m_\perp\sigma^z)\psi_j + \text{H.c.}\right] \\
        &\quad - \frac{1}{2}\sum_j \left[\psi^{\dagger}_{j+\vb{a}_{z}} m_z\sigma^z\psi_j + \text{H.c.}\right]
        + m_0\sum_j \psi^\dagger_j\sigma^z\psi_j,
    \end{aligned}
\end{equation}
where $\psi^\dagger_j = (c^\dagger_{j,s}, c^\dagger_{j,p})$ creates electrons in the $s$- and $p$-orbitals at site $j$. The parameters $m_\perp$ and $m_z$ denote intra-orbital hopping amplitudes in the $xy$-plane and along the $z$-axis, respectively, while $m_0$ is the on-site energy difference between the two orbitals. The inter-orbital hopping is encoded in the terms proportional to $t\sigma^{x}$ and $t\sigma^{y}$, given by Slater-Koster integrals between nearest neighbors.  In equilibrium, this model supports two Dirac points located at $\vb{K}_\pm = (0, 0, \pm k_{z0})$, where the position $k_{z0}$ is determined by $\cos(k_{z0}a) = (m_0 + 2m_\perp)/m_z$.

The modulation of the inter-orbital hopping matrices by optical phonons arises from two primary effects: bond stretching and relative rotation of orbitals \cite{Viscoelastic-response-of-topological-tight-binding,Elastic-Gauge-Fields-in-Weyl,Chiral-Anomaly-from-Strain-Induced}.  As shown in Fig.~\ref{fig1}(b), a phonon mode polarized along the direction $\hat{e}_1$ stretches the bond between neighboring sites, modifying the hopping amplitude in that direction. This leads to a change in the inter-orbital hopping matrix: $
t(u)\sigma^1 \approx \left(t + u\,\partial t/\partial a\right)\sigma^1 = t(1 - \beta u/a)\sigma^1$, where $a$ is the equilibrium bond length and $\beta = -\partial \ln t / \partial \ln a$ is a dimensionless parameter. In contrast, hopping along the orthogonal direction $\hat{e}_2$ remains unaffected.  To express this modulation in the global Cartesian basis, we project the deformed hopping terms along $\hat{x}$ and $\hat{y}$:
\begin{equation}
    \begin{aligned}
        t\sigma^x &\to t\left(1 - \frac{\beta u}{a}\right)\cos\theta\,\sigma^1 - t\sin\theta\,\sigma^2,\\
        t\sigma^y &\to t\left(1 - \frac{\beta u}{a}\right)\sin\theta\,\sigma^2 + t\cos\theta\,\sigma^1.
    \end{aligned}
\end{equation}
Substituting these anisotropically modified hopping matrices into Eq.~(\ref{tight-binding}) and transforming to momentum space, $\sum_{\vb{k}} \psi^\dagger_{\vb{k}}\mathcal{H}(\vb{k})\psi_{\vb{k}}$, we find that the Fermi velocity becomes directionally renormalized near each Dirac point $\vb{K}_\chi$. This anisotropic velocity renormalization corresponds to the frame field $e^{\mu}_A$ introduced in Eq.~(\ref{frame field}).

On the other hand, the emergent gauge field arises from the relative rotation of the $s$ and $p$ orbitals between neighboring sites in the $z$ direction.
To distinguish between the inversion-even and odd optical modes, we formally double the unit cell to include two lattice sites. 
The relative rotation induced by $\vb{u}$ gives rise to a new inter-orbital term along $\hat z$ of the form $it(\beta/a)(u_x \sigma^x + u_y \sigma^y)$.
For an inversion-even mode, this hopping modifies the Hamiltonian by adding a term $\sum_{\vb{k}}\psi^\dagger_{\vb{k}} [t(\beta/a)\sin(k_z a)(u_x \sigma^x + u_y \sigma^y)]\psi_{\vb{k}}$, which contributes to the emergent gauge field $\chi \vb{a} = \chi k_{z0}(\beta/a)\vb{u}$, where $\chi = \pm 1$ is the valley index.

By contrast, this term vanishes identically for inversion-odd phonon modes, resulting in a vanishing emergent gauge field.  This difference is a consequence of symmetry.
In general, the minimal coupling $\vb{k} - \chi \vb{a}$ must be compatible with the little group at the Dirac point.
In our model, the little group at $\vb{K}_\chi$ includes the combined operation $T \times I$, where $T$ denotes time-reversal and $I$ denotes inversion. Since $\vb{k}$ is even under $T \times I$ and the valley index $\chi$ is also even, the phonon displacement $\vb{u}$ must be even under inversion for the coupling to be symmetry-allowed. As a result, inversion-odd modes cannot couple to electrons as a gauge field, while inversion-even modes can.

\textit{Angular momentum channels.}—
To further classify the effects of electron-phonon coupling, we analyze how phonon modes deform the Fermi surface. 
Starting from the effective Hamiltonian in Eq.~(\ref{effective hamiltonian}), we solve the eigenvalue equation and set the energy to zero to obtain the Fermi surface, which is deformed by phonons as
\begin{equation}
\label{Fermi surface}
\begin{aligned}
    \sum_{\mu,\nu} &\delta^{\mu\nu} k_\mu k_\nu = k_F^2 \to\\
    &\sum_{\mu,\nu} g^{\mu\nu}(\vb{u})\left[k_\mu - \chi a_\mu(\vb{u})\right]\left[k_\nu - \chi a_\nu(\vb{u})\right] = k_F^2,
\end{aligned}
\end{equation}
where $k_F = \varepsilon_F/ v_F$ is the Fermi momentum, and $g^{\mu\nu} = e_A^\mu e_B^\nu \delta^{AB}$ is the spatial metric tensor \cite{carroll2004spacetime}.
The angular Fourier components of the Fermi surface deformation provide a classification scheme for the electron-phonon coupling effects: 
the dipolar ($l=1$) channel, which describes a rigid shift of the Fermi surface, corresponds to the emergent gauge field, while the frame field contributes to both a monopolar ($l=0$) deformation [Fig.~\ref{fig1}(c)] and a quadrupolar ($l=2$) distortion. 
The latter leads to a nematic Fermi surface, associated with the traceless part of the spatial metric,
\begin{equation}
\label{quadrupolar}
g^{\mu\nu,(2)}(\vb{u}) \approx -\frac{2\beta u}{a}\left( \hat{e}_1^\mu \hat{e}_1^\nu - \frac12\delta^{\mu\nu} \right),
\end{equation}
where $(\hat{e}_1^x, \hat{e}_1^y) = (\cos\theta, \sin\theta)$ is the phonon polarization direction.
This quadrupolar component shares the same tensorial form as a nematic order parameter, capturing an elliptic distortion of the Fermi surface.
We note that the formation of the nematic Fermi surface is dynamically driven by the phonon excitations, which is distinct from the static nematic order that arises from spontaneous symmetry breaking via Pomeranchuk instability \cite{Nematic-Fermi-Fluids}.

The angular momentum classification also clarifies the origin of the phonon orbital magnetic moment. In the dipolar ($l=1$) channel [Fig.~\ref{fig1}(d)], the phonon shifts the center of the Fermi surface and induces a circular motion around the Dirac point, generating an orbital magnetic moment. In the quadrupolar ($l=2$) channel [Fig.~\ref{fig1}(e)], the phonon distorts the Fermi surface into a nematic shape and drives its rotation, also producing an orbital moment.  These two mechanisms are momentum-space analogs of the two adiabatic contributions to phonon magnetism previously identified in band insulators~\cite{Geometric-orbital-magnetization,Phonon-Magnetic-Moment,Gate-Tunable-Phonon}.

\textit{Phonon magnetic moment.}---
To quantify the two contributions to the phonon magnetic moment, we integrate out the Dirac fermions in the effective Hamiltonian in Eq.~(\ref{effective hamiltonian}) using established low-energy field theories for emergent gauge and frame fields.
In the gauge field channel, the Chern-Simons action $S_{H}[a_\mu] = (\sigma_{xy}/2)\int dt d^x \epsilon^{\mu\nu\rho} a_{\mu}\partial_{\nu}a_{\rho}$ links the phonon moment directly to the electrical Hall conductivity, yielding $\mu_{\text{ph}} =  ({\hbar^3 k_{z0}^2 \beta^2}/{e^2 v_F^2\rho_I a^2}){\sigma_{xy}}/{B}$, where $\rho_I$ is the ion mass density and $B$ is the magnetic field \cite{Gauge-theory-of-giant-phonon}.
(We restore $e$ and $\hbar$ hereafter.)
In this work, we focus on the contribution from the frame field channel. In (2+1)-dimensional Chern insulators, integrating out Dirac fermions coupled to the coframe field $e_\mu^A$ leads to an effective action~\cite{Torsional-Response-and-Dissipationless-Viscosity,Torsional-anomalies}: $S_{H}[e_{\mu}^A] = (\eta_H/2)\int dt d^2x\ \epsilon^{\mu\nu\rho} e_{\mu}^A \partial_{\nu} e_{\rho}^B \delta_{AB}$,
where $\eta_H$ is the Hall viscosity coefficient of the Dirac fluid, $\epsilon^{\mu\nu\rho}$ is the Levi-Civita symbol, and $e_{\mu}^A$ is the coframe field defined as the inverse of $e^{\mu}_A$ satisfying $e_{\mu}^A e_B^{\mu} = \delta^A_B$ \cite{supplemental}. By stacking Chern insulators in momentum space, this term can also be generalized to describe the (3+1)-dimensional Dirac system considered here \cite{Torsional-Response-and-Dissipationless-Viscosity,Burkov2012}.
This action encodes the Hall viscosity response of the Dirac system, a dissipationless transport coefficient that characterizes the transverse stress response to metric perturbations in the presence of broken time-reversal symmetry~\cite{Viscosity-of-Quantum-Hall}. 

In our context, the optical phonon acts as a dynamic metric perturbation, generating a Hall viscosity response in the electron system. 
Expressing $e_\mu^A$ in terms of the phonon displacement $\vb{u}$, we obtain a new term in the phonon effective action, 
$S_{\text{eff}}[\vb{u}] = S_{0}[\vb{u}] + S_H[\vb{u}]$, where $S_{0}[\vb{u}] = \int dt d^2x \ \rho_I \left(\dot{\vb{u}}^2 - \omega_0^2 \vb{u}^2\right)/2$ with $\rho_I$ the ion mass density and $\omega_0$ the bare frequency.
The new term is
\begin{equation}
\label{effective action}
\begin{aligned}
S_H[\vb{u}] &= \frac{\eta_H\beta^2}{2a^2}\int dt d^2x \ \epsilon^{\mu\nu} u_{\mu} \dot u_{\nu},
\end{aligned}
\end{equation} 
This term breaks time-reversal symmetry and leads to a splitting of the chiral phonon modes $u_{\pm} = (u_x \pm i u_y)/\sqrt{2}$. 
Physically, it reflects a feedback mechanism: the phonon motion drags the electron fluid, inducing a Hall viscosity response that in turn exerts a transverse force on the phonon—manifesting as a torque and splitting the phonon frequencies: $\omega_{\pm} = \sqrt{\omega_0^2 + \delta\omega^2} \pm \delta\omega$, where the splitting is $\delta\omega = {\eta_H\beta^2}/{2a^2\rho_I}$ \cite{supplemental}.

\begin{table}
    \caption{\label{table1}Classification of electron-phonon coupling channels by angular momentum and inversion symmetry. The $l=1$ (gauge field) channel is forbidden for inversion-odd modes, while $l=0,2$ (frame field) channels are allowed for both parities. The phonon magnetic moment $\mu_{\text{ph}}$ is proportional to $\sigma_{xy}$ in the gauge field channel and to $\eta_H$ in the quadrupolar frame field channel.}
\begin{ruledtabular}
\begin{tabular}{c c c c}
    & \makebox[0.8in][c]{$l=0$ (Frame)} & 
      \makebox[0.8in][c]{$l=1$ (Gauge)} & 
      \makebox[0.8in][c]{$l=2$ (Frame)} \\
   \hline
   Inversion-even & $\checkmark$ & $\checkmark$ & $\checkmark$ \\
   Inversion-odd  & $\checkmark$ & $\times$ & $\checkmark$ \\
   \hline
   $\mu_{\text{ph}}$ & 0 & $\sigma_{xy}$ & $\eta_H$ \\
   \end{tabular}
\end{ruledtabular}
\end{table}

To estimate the Hall viscosity of the electron fluid, we employ a semiclassical approach. In this regime, $\eta_H$ depends on both the magnetic field $B$ and the electron transport lifetime $\tau$, and is given by $\eta_H = {2\eta_0\,\omega_c\tau}/{(1 + 4\omega_c^2 \tau^2)},$
where $\eta_0 = n_e m^* v_F^2 \tau / 4$ is the zero-field shear viscosity, $\omega_c = eB/m^*$ is the cyclotron frequency, $m^*$ is the effective mass, and $n_e$ is the electron density~\cite{Negative-Magnetoresistance-in-Viscous,Hydrodynamic-Electron-Flow,Nonlocal-transport-and-the-Hall-viscosity}. 

In the weak-field limit ($\omega_c \tau \ll 1$), $\eta_H$ scales linearly with $B$, resulting in a linear splitting of the phonon frequencies: $\hbar \omega_{\pm} = \hbar \omega_0 \pm \mu_{\text{ph}} B$,
which can be interpreted as a phonon Zeeman effect. The corresponding phonon magnetic moment is
\begin{equation}
\label{phonon magnetic moment}
\mu_{\text{ph}} = \frac{\hbar \beta^2}{\rho_I a^2} \cdot \frac{\eta_H}{B},
\end{equation}
where the factor $\beta/a$ quantifies the strength of electron-phonon coupling via the frame field. As $B$ approaches zero, $\eta_H \approx n_e v_F^2 \tau^2 eB / 2$, implying that $\mu_{\text{ph}}$ becomes independent of $B$.

This result establishes a direct and measurable link between the phonon magnetic moment and the Hall viscosity of the electron fluid through frame field coupling—one of the central findings of this work. A summary of the angular momentum channels and their symmetry properties is provided in Table~\ref{table1}.

\begin{figure}
    \centering
    \includegraphics[width=\linewidth]{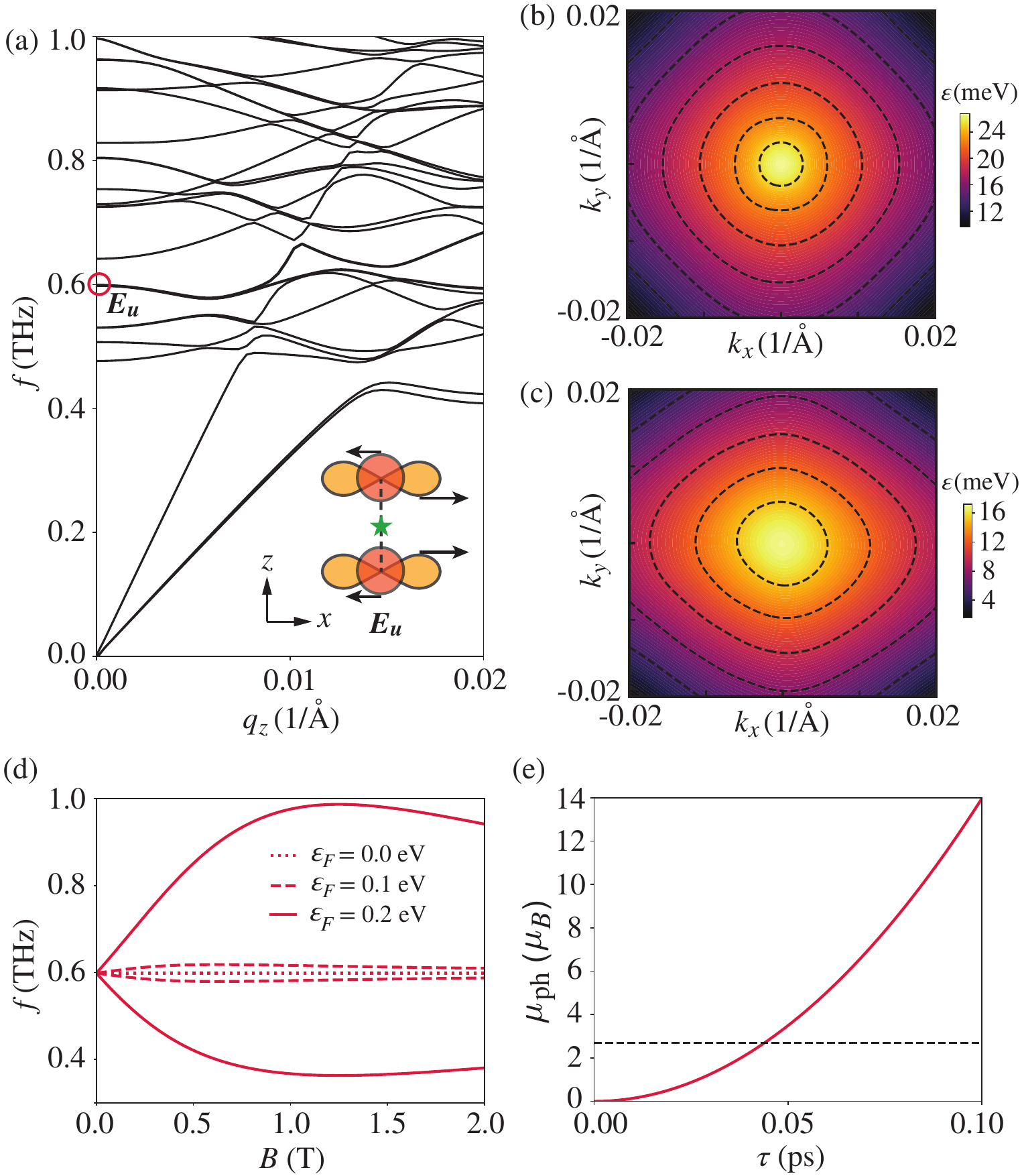}
    \caption{First-principles analysis of the $E_u$ phonon mode in Cd$_3$As$_2$ and its magnetic moment. (a) Phonon spectrum showing the infrared-active $E_u$ mode at 0.6 THz, corresponding to the experimentally measured mode. Inset illustrates its displacement pattern in the $sp$-orbital model. (b) Energy contours of the Dirac cone at $\vb{K}_+$ in equilibrium. (c) Same as (b), but with $E_u$ phonon distortion, showing elliptical deformation consistent with a phonon-induced frame field. (d) Phonon frequency splitting versus magnetic field at $\tau = 0.08$ ps. (e) Phonon magnetic moment versus electron transport lifetime $\tau$ at $\varepsilon_F = 0.1$ eV. Dashed line indicates the experimental value.}
    \label{fig2}
\end{figure}

\textit{Application to Cd$_3$As$_2$.}—
We now apply our theory to the Dirac semimetal Cd$_3$As$_2$, where a large phonon magnetic moment has been observed experimentally \cite{A-Large-Effective}.
Thus far, we have neglected spin degrees of freedom. Including spin introduces two main effects.
First, the $E_u$ phonon mode—by breaking inversion symmetry—can split each Dirac point into a pair of Weyl nodes. However, time-reversal symmetry ensures that the Weyl nodes are displaced in opposite directions in momentum space. As long as the Fermi surface encloses both nodes, their contributions to the phonon magnetic moment cancel, and the frame field mechanism remains the dominant effect.
Second, strong spin-orbit coupling can enable chiral phonons to induce spin polarization, giving rise to a phonon magnetic moment with both spin and orbital components. In this work, we focus exclusively on the orbital contribution and leave spin-related effects for future investigation.

Based on these considerations, we carry out first-principles calculations.
Figure~\ref{fig2}(a) shows the phonon spectrum of Cd$_3$As$_2$ in the THz range. 
The infrared-active $E_u$ optical mode at $0.6$ THz corresponds to the experimentally measured mode.  
Its twofold degeneracy permits the formation of circularly polarized modes. As shown in the inset of Fig.~\ref{fig2}(a), this mode is inversion odd and thus does not generate an emergent gauge field, consistent with our symmetry analysis.

Figures~\ref{fig2}(b) and \ref{fig2}(c) compare the in-plane energy contours of the Dirac cone at the $\vb{K}_+$ valley in equilibrium and under $E_u$-mode distortion, respectively. The elliptic deformation induced by the phonon displacement confirms the emergence of a frame field and the formation of a nematic Fermi surface. Moreover, the absence of a Dirac point shift reinforces the conclusion that the emergent gauge field vanishes for this mode. These results from first-principles calculations are fully consistent with our theoretical predictions. From the slope of the Dirac cone and the averaged phonon eigenvector, we extract an estimate for the electron-phonon coupling parameter: $\beta/a \approx 632\text{ \AA}^{-1}$~\cite{supplemental}.

For numerical estimates, we adopt the following parameters for Cd$_3$As$_2$: $v_F \approx 10^6$ m/s and $\rho_I \approx 3.03 \times 10^3$ kg/m$^3$. Figure~\ref{fig2}(d) shows the calculated phonon frequency splitting as a function of magnetic field at a fixed transport lifetime $\tau = 0.08$ ps. In the weak-field limit, the splitting is linear in $B$, while in the strong-field limit, it becomes inversely proportional to $B$. The asymmetric evolution of the $\omega_+$ and $\omega_-$ branches reproduces key features of the experimental data~\cite{A-Large-Effective}.
Finally, in Fig.~\ref{fig2}(e), we plot the phonon magnetic moment as a function of $\tau$ at fixed Fermi energy $\varepsilon_F = 0.1$ eV. For the experimental sample, $\tau \sim 0.1$ ps, and the calculated phonon magnetic moment is in good agreement with the measured value. This supports the conclusion that the frame field mechanism developed here accounts for the observed phonon magnetic moment in Cd$_3$As$_2$.

\textit{Conclusion and discussion.}—
We have presented a general mechanism by which electron-phonon coupling can give rise to sizable phonon magnetic moments in Dirac materials. By analyzing the angular momentum components of phonon-induced Fermi surface deformations, we classify the phonon field as coupling to electrons via either an emergent gauge field or a frame field. In the gauge (frame) field channel, the resulting phonon magnetic moment is proportional to the electronic Hall conductivity (viscosity).

We applied the frame field mechanism to Cd$_3$As$_2$ and found quantitative agreement between the calculated phonon magnetic moment and experimental measurements, using realistic material parameters. Beyond explaining the observed phonon moment, our results point to a new avenue for probing the elusive Hall viscosity of electron fluids through phonon-based measurements—particularly in systems where competing effects, such as spin-phonon coupling, can be minimized or systematically disentangled.

\begin{acknowledgements}
We thank Alexander Balatsky for the helpful discussion. The work at UW was supported by DOE Award No. DE-SC0012509. The work at LANL was carried out under the auspices of the U.S. DOE NNSA under contract No. 89233218CNA000001 through the LDRD Program, and was supported by the Center for Nonlinear Studies at LANL, and was performed, in part, at the Center for Integrated Nanotechnologies, an Office of Science User Facility operated for the U.S. DOE Office of Science, under user proposals $\#2018BU0010$ and $\#2018BU0083$.
\end{acknowledgements}

%

\pagebreak
\widetext
\begin{center}
\textbf{\large Supplemental Information for ``Geometric Origin of Phonon Magnetic Moment in Dirac Materials''}
\end{center}
\setcounter{equation}{0}
\setcounter{figure}{0}
\setcounter{table}{0}
\setcounter{page}{1}
\makeatletter
\renewcommand{\theequation}{S\arabic{equation}}
\renewcommand{\thefigure}{S\arabic{figure}}
\renewcommand{\citenumfont}[1]{S#1}

\section{Dirac fermions in curved space}
We briefly review the theory of Dirac fermions in curved space \cite{Fock1929,Weyl1929,Gauge-fields-in-graphene}.
In flat space, the Lagrangian density for Dirac fermions is given by
$\mathcal{L}_D = \psi^\dagger \gamma^0 (i\gamma^A \partial_A - m) \psi$,
where $\gamma^A$ are the gamma matrices, $m$ is the fermion mass, and $\psi$ is the Dirac spinor.
In curved space, the Lagrangian is modified to incorporate the frame field and spin connection,
\begin{equation}
\mathcal{L}_D = \det(\vb{e}) \psi^\dagger \gamma^0 \left[i\gamma^A e_A^{\mu}(x) D_\mu - m\right] \psi,
\end{equation}
where $e_A^{\mu}(x)$ is the frame field (or vielbein), $D_\mu = \partial_\mu + \Gamma_\mu$ is the covariant derivative, and $\Gamma_\mu$ is the geometric connection.
The geometric connection is given by
$\Gamma_\mu = \frac{1}{2} \omega_{AB\mu} \Sigma^{AB}$,
where $\omega_{AB\mu}$ is the spin connection, and $\Sigma^{AB} = \frac{1}{4}[\gamma^A,\gamma^B]$ are the generators of the Lorentz group.
The spin connection $\omega_{AB\mu}$ depends on the derivatives of the frame field $e_A^{\mu}(x)$ and ensures local Lorentz invariance in curved space.

The frame field $e_A^{\mu}(x)$ defines the local orthonormal basis vectors $\vb{e}_A$ in curved space by relating them to the global coordinate basis vectors $\hat{e}_\mu = \partial_{\mu}$ \cite{carroll2004spacetime}:
\begin{equation}
\vb{e}_A = e_A^{\mu}(x)\hat{e}_\mu.
\end{equation}
By construction, the $\vb{e}_A$ are orthonormal, such that their inner product satisfies $g(\vb{e}_A,\vb{e}_B) = \eta_{AB}$, which can be equivalently expressed as
\begin{equation}
g^{\mu\nu} = e_A^{\mu} e_B^{\nu} \eta^{AB},
\end{equation}
where $g^{\mu\nu}$ is the metric tensor of the curved space and $\eta^{AB}$ is the Minkowski metric.
Thus, the frame field $e_A^{\mu}(x)$ can be interpreted as the “square root” of the metric tensor.

The canonical momentum conjugate to the Dirac field $\psi$ is given by
\begin{equation}
\Pi = \frac{\partial \mathcal{L}_D}{\partial(D_0\psi)} = \det(\vb{e})\, i\psi^\dagger \gamma^0 e^0_{A}(x)\gamma^A.
\end{equation}
Using this, the Hamiltonian density can be written as
\begin{equation}
\mathcal{H}_D = \Pi D_0\psi - \mathcal{L}_D = \det(\vb{e}) \psi^\dagger \left[\qty(p_{\mu} - i\Gamma_\mu) e_A^{\mu}(x) \alpha^A + m\beta\right].
\end{equation}
For simplicity, we now restrict the global coordinates $\mu$ to two spatial dimensions $(x,y)$, and we choose the representation $\alpha^1 = \sigma^1$, $\alpha^2 = \sigma^2$, and $\beta = \sigma^3$.
In this case, the Dirac Hamiltonian simplifies to
\begin{equation}
H_D = \qty(p_{\mu} - i\Gamma_\mu) e_A^{\mu}(x) \sigma^A + m\sigma^3.
\end{equation}
Compared to the flat-space Dirac Hamiltonian,
$H_D = p_{\mu}\sigma^\mu + m\sigma^3$,
the geometric deformations in curved space are encoded in the frame field $e_A^{\mu}(x)$ and the geometric connection $\Gamma_\mu$, which itself is determined by the frame field.

Importantly, the geometric connection $\Gamma_\mu$ couples to the Dirac fermions in the same form as a gauge field $A_\mu$.
In flat space, the gauge field enters the Dirac Hamiltonian as an explicit additional term $e \vb{A} \cdot \boldsymbol{\sigma}$.
In curved space, however, the gauge field can be absorbed into the geometric connection by redefining
$\Gamma^{\prime}_\mu = \Gamma_\mu + i e A_\mu$.
Thus, the Dirac Hamiltonian can be rewritten as
\begin{equation}
\mathcal{H}_D = \qty(p_{\mu} - i\Gamma_\mu) e_A^{\mu}(x) \sigma^A + m\sigma^3.
\end{equation}
This shows that in the curved space formulation, gauge fields can be geometrized via a modified geometric connection.

In condensed matter systems, lattice distortions are the primary sources of emergent frame fields and spin connections, which provide a natural geometric framework to describe the coupling between lattice degrees of freedom and low-energy electronic excitations.
These distortions can arise from both static and dynamic lattice phenomena.

One example involves topological defects in the crystal lattice, such as dislocations and disclinations \cite{Torsional-anomalies}.
Disclinations, such as an isolated pentagon or heptagon ring in graphene, act as monopole sources of curvature in the effective geometry, giving rise to a nonzero spin connection felt by the electrons.
Dislocations, which introduce a missing or extra half-plane of atoms, generate torsion fields by effectively translating the trajectory of electrons encircling the dislocation core.
This torsion is mathematically described as the field strength associated with the frame field, reflecting the nontrivial topology of the lattice defect.
Beyond defects, smoothly varying lattice distortions such as elastic strain can also generate effective frame fields and spin connections \cite{Torsional-anomalies}.
Strain gradients modify the local metric and thus couple into the low-energy Dirac Hamiltonian, allowing for an elegant geometric description of strain-engineered electronic properties.

In this work, we focus on a dynamic and particularly relevant source of lattice distortion—phonons.
Specifically, we consider optical phonon modes, which involve relative motions of atoms within the unit cell.
In the long-wavelength limit, the distortions induced by such phonons effectively act as a uniform frame field, modifying the local geometry experienced by the electrons, while the corresponding spin connection remains negligible due to the absence of lattice curvature or torsion---like stretching a membrane without bending it.

\section{Tight-binding model}
In this section, we explicitly demonstrate how optical phonons can induce emergent gauge and frame fields coupled to Dirac fermions using a tight-binding model.
We start from the $\vb{k}\cdot\vb{p}$ model for Cd$_3$As$_2$, which captures the essential band inversion between Cd-$5s$ and As-$4p$ orbitals near the $\Gamma$ point~\cite{Three-dimensional-Dirac-semimetal,Chiral-Anomaly-from-Strain-Induced}.
The basis consists of $\ket{S_{\frac{1}{2}},\frac{1}{2}}, \ket{P_{\frac{3}{2}},\frac{3}{2}}, \ket{S_{\frac{1}{2}},-\frac{1}{2}}, \ket{P_{\frac{3}{2}},-\frac{3}{2}}$, and the Hamiltonian is given by
\begin{equation}
\mathcal{H}_{\Gamma}(\vb{k}) = \varepsilon_0(\vb{k}) + M(\vb{k})\sigma^z + A k_x \tau^z \sigma^x + A k_y\sigma^y,
\end{equation}
where $\varepsilon_0(\vb{k}) = C_0 + C_1 k^2_z + C_2 (k^2_x + k^2_y)$ and $M(\vb{k}) = M_0 + M_1 k^2_z + M_2 (k^2_x + k^2_y)$, with $C_{0,1,2}$ and $M_{0,1,2}$ the model parameters, and $\boldsymbol{\sigma}$ and $\boldsymbol{\tau}$ are Pauli matrices in the orbital and spin subspaces, respectively.

Due to the block-diagonal structure of the Hamiltonian, where spin-up and spin-down sectors are decoupled, we focus on the spin-up block formed by $\ket{S_{\frac{1}{2}},\frac{1}{2}}$ and $\ket{P_{\frac{3}{2}},\frac{3}{2}}$.
By employing the standard lattice regularization:
\begin{equation}
\begin{aligned}
k_{x,y} &\to \frac{1}{a}\sin(k_{x,y} a),\\
k_{x,y,z}^2 &\to \frac{2}{a^2}[1-\cos(k_{x,y,z} a)],
\end{aligned}
\end{equation}
we arrive at a tight-binding model on a cubic lattice with lattice constant $a$.
The resulting lattice Hamiltonian is:
\begin{equation}
\mathcal{H}(\vb{k}) = [m_0 + m_z\cos(k_z a) + m_\perp\cos(k_x a) + m_\perp\cos(k_y a)]\sigma^z + t\sin(k_x a)\sigma^x + t\sin(k_y a)\sigma^y,
\end{equation}
where $m_0 = M_0 + 2M_1/a^2 + 4M_2/a^2$ is the on-site energy difference between the two orbitals, $m_z = -2M_1/a^2$ and $m_\perp = -2M_2/a^2$ are intra-orbital hopping amplitudes along the $z$ direction and in the $xy$ plane, respectively, and $t = A/a$ is the inter-orbital hopping amplitude.

This model features two spinless Dirac points at $\vb{K}_{\chi} = (0,0,\chi k_{z0})$, where $\chi = \pm1$ labels the two valleys, and $\cos(k_{z0} a) = (m_0 + 2m_\perp)/m_z$ determines their location.
Expanding the Hamiltonian around $\vb{K}_{\chi}$ yields:
\begin{equation}
\mathcal{H}_{\chi}(\vb{k}) = \hbar v_F \qty(k_x \sigma^x + k_y \sigma^y) + \chi\hbar v_z k_z \sigma^z,
\end{equation}
where $v_F = t a/\hbar$ and $v_z = a m_z\sin(k_{z0} a)/\hbar$.
Despite the complexity of the actual Cd$_3$As$_2$ crystal structure, which has 80 atoms per unit cell, the essential low-energy physics can be captured by this minimal model with $s$- and $p$-orbitals placed on a cubic lattice.

We now transform the Hamiltonian $H = \sum_{\vb{k}}\psi^\dagger_{\vb{k}}\mathcal{H}(\vb{k})\psi_{\vb{k}}$ into real space, yielding:
\begin{equation}
    \label{real_space_H}
    \begin{aligned}
        H &= \frac{1}{2}\sum_j \left[\psi^{\dagger}_{j+\vb{a}_{x}}(it\sigma^{x} - m_\perp\sigma^z)\psi_j + h.c.\right]\\
        &+ \frac{1}{2}\sum_j \left[\psi^{\dagger}_{j+\vb{a}_{y}}(it\sigma^{y} - m_\perp\sigma^z)\psi_j + h.c.\right]\\
        &+ \frac{1}{2}\sum_j \left[\psi^{\dagger}_{j+\vb{a}_{z}}(- m_z\sigma^z)\psi_j + h.c.\right]
        + m_0\sum_j \psi^\dagger_j\sigma^z\psi_j.
    \end{aligned}
\end{equation}
Here, $\psi_j^\dagger = (c_{j,s}^\dagger, c_{j,p}^\dagger)$, and $\vb{a}_{x,y,z}$ are the lattice vectors along each Cartesian direction.

In order to distinguish between the in-plane inversion-odd ($E_u$) and inversion-even ($E_g$) optical phonon modes, we formally double the unit cell along the $z$ direction to include two sites per unit cell. 
Phonon-induced lattice distortions affect the tight-binding model by modifying the hopping amplitudes.
We focus on the modifications to the inter-orbital hoppings between the nearest neighbors, given by two-center Slater-Koster integrals $t_{ij} = \int d^3 r \ \psi^*_\alpha(\vb{r}-\vb{R}_i) H \psi_\beta(\vb{r} - \vb{R}_j)$, where $\vb{R}_j$ is the position of the $j$-th size, and $\psi_{\alpha}$ is the orbital wave function ($\alpha \neq \beta$). 
If the lattice is distorted, then the site positions $\vb{R}_j$ are replaced by $\vb{R}_j + \vb{u}_j$, where $\vb{u}_j$ is the displacement.
Under the frozen phonon approximation and long-wavelength aprroximation, we can treat $\vb{u}_j$ as time-independent and site-independent. 

Expanding $t_{ij}(\vb{u})$ to first order in $\vb{u}$, the modulations of the inter-orbital hopping matrices by optical phonons arises from two primary effects: bond stretching and relative rotation of the two orbitals \cite{Viscoelastic-response-of-topological-tight-binding,Elastic-Gauge-Fields-in-Weyl,Chiral-Anomaly-from-Strain-Induced}.

(1) Bond stretching. 
A phonon mode polarized along the direction $\hat e_1$ stretches the bond length between the neighboring sites, modifying the hopping amplitude in that direction.
This leads to a change in the inter-orbital hopping matrix,
\begin{equation}
    t(\vb{u})\sigma^1 \approx \qty(t + u\frac{\partial t}{\partial a})\sigma^1 = t\qty(1-\frac{\beta u}{a})\sigma^1,
\end{equation}
where $a$ is the equilibrium bond length, and $\beta = -\partial \ln t /\ln a$ is the Grüneisen parameter that characterizes the strength of the bond stretching. 
In contrast, $t\sigma^2$---hopping along the orthogonal direction $\hat e_2$---is not affected by the bond-length change.
If the phonon is polarized in arbitary direction not just along $x$- and $y$-axis, aside from the bond-length change, we also have the bond-angle change contribution. 
Since the phonon polarization is not neccessarily aligned with $x$- or $y$-axis, this contribution can be seen by projecting the deformed hopping matrices into the global Cartesian coordinates.
Then the inter-orbital hopping matrices are modified as
\begin{equation}
    \begin{aligned}
        t\sigma^x &\to t\qty(1-\frac{\beta u}{a}) \cos{\theta}\,\sigma^1 - t\sin{\theta}\,\sigma^2,\\
        t\sigma^y &\to t\qty(1-\frac{\beta u}{a}) \sin{\theta}\,\sigma^1 + t\cos{\theta}\,\sigma^2,
    \end{aligned}
\end{equation}
where $\theta$ is the instantaneous angle between the displacement vector $\vb{u}$ and the $x$ axis.
When the magnitude of the phonon displacement $u$ is zero, this modification is equivalent to a simple rotation of the Pauli matrices: $\sigma^x \to \cos{\theta}\,\sigma^1 - \sin{\theta}\,\sigma^2$ and $\sigma^y \to \sin{\theta}\,\sigma^1 + \cos{\theta}\,\sigma^2$.
The bond stretching induced by nonzero phonon displacement $u$ modifies the hopping matrices anistropically.

(2) Relative rotation of the two orbitals, which only modifies the hopping amplitude between different types of orbitals. 
For our purpose of deriving the effects on the Dirac cones lying on the $z$-axis, the relevant modication is the relative rotation of the $s$ and $p$ orbitals between neighboring sites in the $z$ direction. 
This relative rotation gives rise to a new inter-orbital hopping term along the $z$ direction, which can be expressed as
\begin{equation}
    \begin{aligned}
        t_{ij,z}(\vb{u}) &\approx i\frac{\hat y\cdot (a\hat z\times u_x \hat x)}{a}\frac{\partial t}{\partial a} \sigma^x + i\frac{\hat x\cdot (u_y \hat y \times a\hat z)}{a}\frac{\partial t}{\partial a} \sigma^y,\\
        &= it\frac{\beta u_x}{a}\sigma^x + it\frac{\beta u_y}{a}\sigma^y.
    \end{aligned}    
\end{equation}

Inserting both the bond-stretching and orbital-rotation modifications into the original lattice Hamiltonian Eq.~(\ref{real_space_H}), we obtain the deformed Hamiltonian under a phonon displacement field $\vb{u}$:
\begin{equation}
    \begin{aligned}
        H(\vb{u}) &= \frac{1}{2}\sum_j \psi^{\dagger}_{j+\vb{a}_{x}}\left[it\qty(1-\frac{\beta u}{a})\cos\theta\,\sigma^{1} -it\sin\theta\,\sigma^2 - m_\perp\sigma^z\right]\psi_j + h.c.\\
        &+ \frac{1}{2}\sum_j \psi^{\dagger}_{j+\vb{a}_{y}}\left[it\qty(1-\frac{\beta u}{a})\sin\theta\,\sigma^{1} + it\sin\theta\,\sigma^2 - m_\perp\sigma^z\right]\psi_j + h.c.\\
        &+ \frac{1}{2}\sum_j \psi^{\dagger}_{j+\vb{a}_{z}}\qty(it\frac{\beta u_x}{a}\sigma^x + it\frac{\beta u_y}{a}\sigma^y)\psi_j \pm \psi^{\dagger}_{j-\vb{a}_{z}}\qty(it\frac{\beta u_x}{a}\sigma^x + it\frac{\beta u_y}{a}\sigma^y)\psi_j +h.c.\\
        &+ \frac{1}{2}\sum_j \psi^{\dagger}_{j+\vb{a}_{z}}(- m_z\sigma^z)\psi_j + h.c.
        + m_0\sum_j c^\dagger_j\sigma^zc_j,
    \end{aligned}
\end{equation}
where the $\pm$ sign in the third line distinguishes the inversion-odd $E_u$ and inversion-even $E_g$ phonon modes, respectively. 
We now transform the Hamiltonian into momentum space.
For the inversion-even $E_g$ mode, the resulting Hamiltonian reads:
\begin{equation}
    \begin{aligned}
        H_{E_g}(\vb{u}) &= \sum_{\vb{k}} \psi^{\dagger}_{\vb{k}}\left[t\sin(k_x a)\qty(1-\frac{\beta u}{a})\cos\theta\,\sigma^{1} -t\sin(k_x a)\sin\theta\,\sigma^2 \right]\psi_{\vb{k}}\\
        &+ \sum_{\vb{k}} \psi^{\dagger}_{\vb{k}}\left[t\sin(k_y a)\qty(1-\frac{\beta u}{a})\sin\theta\,\sigma^{1} + t\sin(k_y a)\cos\theta\,\sigma^2 \right]\psi_{\vb{k}}\\
        &+ \sum_{\vb{k}} \psi^{\dagger}_{\vb{k}}\left[t\sin(k_z a)\frac{\beta}{a}(u_x\sigma^x + u_y\sigma^y)\right]\psi_{\vb{k}}\\
        &+ \sum_{\vb{k}} \psi^{\dagger}_{\vb{k}}\left[ m_\perp \cos(k_x a) + m_\perp \cos(k_y a)  + m_z\cos(k_z a) + m_0\right]\sigma^z\psi_{\vb{k}}.
    \end{aligned}
\end{equation}
Expanding the above Hamiltonian near the Dirac point $\vb{K}_\chi$, we arrive at the continuum limit:
\begin{equation}
    \begin{aligned}
        \mathcal{H}_{E_g} (\vb{k},\vb{u}) &= \hbar v_F k_x \qty(1-\frac{\beta u}{a})\cos\theta\,\sigma^{1} + \hbar v_F k_x (-\sin\theta)\,\sigma^2\\
        & + \hbar v_F k_y \qty(1-\frac{\beta u}{a})\sin\theta\,\sigma^1+ \hbar v_F k_y \cos\theta\,\sigma^2\\
        & - \chi t\beta k_{z0}(u_x\sigma^x + u_y\sigma^y) + \chi\hbar v_z k_z \sigma^z,
    \end{aligned}
\end{equation}
For the inversion-odd $E_u$ mode, the relative rotation of the $s$ and $p$ orbitals along the $z$ direction results in the term:
\begin{equation}
    \frac{1}{2}\sum_j \psi^{\dagger}_{j+\vb{a}_{z}}\qty(it\frac{\beta u_x}{a}\sigma^x + it\frac{\beta u_y}{a}\sigma^y)\psi_j + \psi^{\dagger}_{j-\vb{a}_{z}}\qty(it\frac{\beta u_x}{a}\sigma^x + it\frac{\beta u_y}{a}\sigma^y)\psi_j +h.c.,
\end{equation}
which cancels exactly in momentum space.
This results in a vanishing emergent gauge field for the $E_u$ mode.
The resulting momentum-space Hamiltonian for the $E_u$ mode is therefore:
\begin{equation}
    \begin{aligned}
        H_{E_u}(\vb{k},\vb{u}) &= \sum_{\vb{k}} \psi^{\dagger}_{\vb{k}}\left[t\sin(k_x a)\qty(1-\frac{\beta u}{a})\cos\theta\,\sigma^{1} -t\sin(k_x a)\sin\theta\,\sigma^2 \right]\psi_{\vb{k}}\\
        &+ \sum_{\vb{k}} \psi^{\dagger}_{\vb{k}}\left[t\sin(k_y a)\qty(1-\frac{\beta u}{a})\sin\theta\,\sigma^{1} + t\sin(k_y a)\cos\theta\,\sigma^2 \right]\psi_{\vb{k}}\\
        &+ \sum_{\vb{k}} \psi^{\dagger}_{\vb{k}}\left[ m_\perp \cos(k_x a)  + m_\perp \cos(k_y a) + m_z\cos(k_z a) + m_0\right]\sigma^z\psi_{\vb{k}}.
    \end{aligned}
\end{equation}
And its continuum limit becomes
\begin{equation}
    \begin{aligned}
        \mathcal{H}_{E_u} (\vb{k},\vb{u}) &= \hbar v_F k_x \qty(1-\frac{\beta u}{a})\cos\theta\,\sigma^{1} + \hbar v_F k_x (-\sin\theta)\,\sigma^2\\
        & + \hbar v_F k_y \qty(1-\frac{\beta u}{a})\sin\theta\,\sigma^1+ \hbar v_F k_y \cos\theta\,\sigma^2\\
        & + \chi\hbar v_z k_z\,\sigma^3.
    \end{aligned}
\end{equation}
We compare the two effective Hamiltonians to the general Dirac Hamiltonian in curved space
\begin{equation}
    \mathcal{H} = v_F(p_\mu - e A_\mu - i\Gamma_\mu)e_A^{\mu}\sigma^A,
\end{equation}
where $A_\mu$ is the gauge field, $e_A^{\mu}$ is the frame field (vierbein), and $\Gamma_\mu$ involves the spin connection.
We find that the spin connection is zero, and both modes exhibit a frame field $e_A^{\mu}$, given by
\begin{equation}
    e_A^{\mu} = \begin{bmatrix}
        \qty(1-\frac{\beta u}{a})\cos\theta & \qty(1-\frac{\beta u}{a})\sin\theta\\
         -\sin\theta & \cos\theta
    \end{bmatrix}.
\end{equation}
In addition, only the $E_g$ mode supports an emergent gauge field:
\begin{equation}
    \chi a_x = \chi\frac{ t\beta k_{z0}}{v_F} u_x, \quad \chi a_y = \chi\frac{ t\beta k_{z0}}{v_F} u_y.
\end{equation}
This contrast between the two modes is a consequence of symmetry.
In general, the emergent gauge field must be compatible with the little group at the Dirac point $\vb{K}_\chi$.
In our model, the little group consists of $C_{4z}$ and the combined operation $T\times I$, where $C_{4z}$ is a fourfold rotation around the $z$ axis, $T$ is time-reversal, and $I$ is inversion.
Both the electron momentum $\vb{k}$ and the phonon displacement $\vb{u}$ transform identically under $C_{4z}$.
Under $T\times I$, $\hbar \vb{k}$ is even, and the valley index $\chi$ is also even.
Therefore, for the combination $\chi \vb{u}$ to couple as a gauge field, $\vb{u}$ must be inversion even.
This symmetry requirement prohibits the inversion-odd $E_u$ mode from coupling as an emergent gauge field, while allowing the inversion-even $E_g$ mode.

\section{Electron Hall viscosity}
\subsection{Integrating out the fermions}
To quantitatively formulate how the Hall viscosity of the Dirac fluid modifies the phonon dynamics, we integrate out the Dirac fermions in the presence of a frame field using low-energy field theories.
Since the frame field $e_A^{\mu}$ is dimensionless, the Hall viscosity coefficient $\eta_H$ must be scaled by $1/[\text{length}]^2$.
A closely related problem has been extensively studied in the context of the quantum Hall effect, where the Hall viscosity arises from integrating out electrons in a magnetic field~\cite{Viscosity-of-Quantum-Hall}.
In that case, the Hall viscosity is found to be quantized in units of ${\hbar}/{8\pi l_B^2}$, with the characteristic length scale set by the magnetic length $l_B$.

More relevant to our case is the Dirac fermion coupled to a static frame field.
Refs.~\cite{Torsional-Response-and-Dissipationless-Viscosity, Torsional-anomalies} showed that, by integrating out massive Dirac fermions in a Chern insulator, an effective topological action for the frame field is obtained:
\begin{equation}
S_{H}[e_{\mu}^A] = \frac{\eta_H}{2}\int d^3x, \epsilon^{\mu\nu\rho}e_{\mu}^A \partial_\nu  e_{\rho}^B \delta_{AB},
\end{equation}
where $\epsilon^{\mu\nu\rho}$ is the Levi-Civita symbol and $\delta_{AB}$ is the Kronecker delta in the local tangent space.
In this case, the Hall viscosity is given by $\eta_H = \hbar/(8\pi l_m^2)$, where $l_m = \hbar v_F/m$ is the length scale associated with the time-reversal symmetry breaking Dirac mass $m$.
By stacking Chern insulators in momentum space, this effective action can also be generalized to describe the (3+1)-dimensional Dirac system considered here \cite{Burkov2012}.

Additionally, the Hall viscosity for massless Dirac fermions in the presence of a magnetic field and geometric deformations has been studied in Ref.~\cite{Hall-viscosity-and-momentum-transport}.
In this setup, the contribution to the Hall viscosity from each Landau level is found to be:
\begin{equation}
\eta_H^{(0)} = \frac{\hbar}{8\pi l_B^2},\quad \text{and}\quad \eta_H^{(n\ne0)} = \frac{\hbar |n|}{4\pi l_B^2}.
\end{equation}
Summing over the filled Landau levels up to a filling fraction $\nu = \hbar n_e / eB$, the total Hall viscosity becomes:
\begin{equation}
\eta_H = \frac{\hbar \nu}{8\pi l_B^2}.
\end{equation}

These results describe the Hall viscosity in the strong-field (quantum Hall) regime, where the Landau levels are well-resolved.
They provide the foundation for our subsequent semiclassical analysis in the weak-field limit.

\subsection{Semiclassical limit of Hall viscosity}
In the semiclassical regime, where the magnetic field is weak and the Landau level filling fraction is large, the Hall viscosity can be calculated using the Boltzmann transport equation~\cite{Negative-Magnetoresistance-in-Viscous,Hydrodynamic-Electron-Flow,Nonlocal-transport-and-the-Hall-viscosity}.
Hall viscosity characterizes the antisymmetric part of the stress response of the electron fluid when time-reversal symmetry is broken.
The stress tensor in such a fluid can be expressed as $-P\delta_{ij} + T_{ij}$, where $P$ is the static pressure and $T_{ij}$ is the shear stress tensor~\cite{landau1987fluid}.
Within the semiclassical framework, the shear stress tensor can be written as
\begin{equation}
    \begin{aligned}
        \vb{T} = \mathcal{N}_0 p_F v_F&\int \frac{d\phi_p}{2\pi} \mathcal{F}(\vb{r},\phi_p,t)\begin{pmatrix}
            \cos^2\phi_p & \cos\phi_p\sin\phi_p\\
            \sin\phi_p\cos\phi_p & \sin^2 \phi_p    
        \end{pmatrix},
    \end{aligned}
\end{equation}
where $\mathcal{N}_0$ is the density of states at the Fermi energy, $p_F$ is the Fermi momentum, and $\phi_p$ is the angular coordinate of the electron momentum $\vb{p}$.
Here, $\mathcal{F}(\vb{r},\phi_p,t)$ describes the deviation from the equilibrium distribution function $f_0(\varepsilon)$ and is defined via
\begin{equation}
f(\vb{r},\vb{p},t) = f_0(\varepsilon) - \frac{\partial f_0}{\partial \varepsilon}\mathcal{F}(\vb{r},\phi_p,t).
\end{equation}
The evolution of $\mathcal{F}(\vb{r},\phi_p,t)$ is governed by the linearized Boltzmann equation,
\begin{equation}
\frac{\partial \mathcal{F}}{\partial t} + \vb{v}_p\cdot \nabla \mathcal{F} + \omega_c\frac{\partial \mathcal{F}}{\partial \phi_p} = \mathcal{I}[\mathcal{F}],
\end{equation}
where $\vb{v}_p = \nabla_p \varepsilon$ is the electron velocity, $\omega_c = eB/m^*$ is the cyclotron frequency arising from the external magnetic field $B$, and $\mathcal{I}[\mathcal{F}]$ denotes the collision integral.

To analyze the angular momentum channels of Fermi surface deformation, we expand $\mathcal{F}(\vb{r},\phi_p,t)$ in angular harmonics:
\begin{equation}
\mathcal{F}(\vb{r},\phi_p,t) = \sum_{l=-\infty}^{+\infty} e^{-il\phi_p}\mathcal{F}_l(\vb{r},t).
\end{equation}
Expressed in terms of these components, the shear stress tensor becomes
\begin{equation}
\begin{aligned}
\vb{T} =&\frac{\mathcal{N}_0 p_F v_F}{4}\begin{pmatrix}
2\mathcal{F}_{0} + \mathcal{F}_{2} + \mathcal{F}_{-2} & i\mathcal{F}_{2} -i \mathcal{F}_{-2}\\
-i\mathcal{F}{2} + i\mathcal{F}{-2}&  2\mathcal{F}{0} -\mathcal{F}{2} - \mathcal{F}{-2}
\end{pmatrix}.
\end{aligned}
\end{equation}

In the hydrodynamic regime, the collision integral can be approximated in linearized form.
By truncating the expansion at $|l| = 2$ and assuming a relaxation time approximation, we obtain
\begin{equation}
    \begin{aligned}
        \mathcal{F}_{\pm 2} = -\frac{v_F}{2}\frac{(\partial_x \mp i\partial_y)\mathcal{F}_{\pm1}(\vb{r})}{1/\tau \pm 2i\omega_c},
    \end{aligned}
\end{equation}
where $\tau$ is the transport lifetime.
Thus, the total shear stress can be expressed as:
\begin{equation}
\vb{T} = \frac{\mathcal{B}}{\bar{n}}n(\vb{r}) + \sigma'(\vb{r}),
\end{equation}
where $\mathcal{B}$ is the bulk modulus, $\bar{n}$ is the average density, and $\sigma'$ is the traceless shear stress given by
\begin{equation}
    \begin{aligned}
        \bm{\sigma}' =& m^*\nu_s\begin{pmatrix}
        \partial_x J_x - \partial_y J_y & \partial_x J_y + \partial_y J_x\\
        \partial_x J_y + \partial_y J_x & -\partial_x J_x + \partial_y J_y
    \end{pmatrix}\\
    &+ m^* \nu_H\begin{pmatrix}
        \partial_x J_y + \partial_y J_x & -\partial_x J_x + \partial_y J_y\\
        \partial_x J_x - \partial_y J_y & --\partial_x J_y - \partial_y J_x
    \end{pmatrix},
    \end{aligned}
\end{equation}
where $\vb{J}$ is the electron current density.
The kinetic (dissipative) shear viscosity is given by
\begin{equation}
    \nu_s = \frac{v^2_F}{4}\frac{\tau}{1+ 4\omega_c^2\tau^2},
\end{equation}
and the kinetic Hall (nondissipative) viscosity is
\begin{equation}
    \nu_H = \frac{v_F^2}{2}\frac{\omega_c\tau^2}{1+4\omega_c^2\tau^2}.
\end{equation}
Finally, the Hall viscosity is related to the kinetic Hall viscosity $\nu_H$ by
\begin{equation}
    \eta_H = n_e m^* \nu_H
\end{equation}
By its definition, the Hall viscosity quantifies the shear stress response of the electron fluid to the quadrupolar (elliptic) deformation of the Fermi surface—captured by the $l = \pm 2$ harmonic component of $\mathcal{F}$.
Phonons, when they couple to electrons as metric perturbations or frame fields in the low-energy theory, induce precisely such elliptic deformations of the Fermi surface and thus generate a Hall viscosity response in the electron fluid.

\subsection{Nematic Fermi surface}

To show how the phonon-induced frame field deforms the electron fluid into a nematic Fermi surface, we consider the following effective Hamiltonian describing Dirac fermions coupled to a general frame field in two dimensions
\begin{equation}
\begin{aligned}
\mathcal{H} &= \hbar v_F (k_x e_1^x + k_y e_1^y)\sigma^1 + \hbar v_F (k_x e_2^x + k_y e_2^y)\sigma^2 - \varepsilon_F\\
&= h_1\sigma^1 + h_2\sigma^2 - \varepsilon_F.
\end{aligned}
\end{equation}
Here, $e_A^\mu$ represents the frame field components that encode the local deformation of the electron system, while $\sigma^{1,2}$ are Pauli matrices acting in the orbital space.
The resulting energy dispersion is given by
\begin{equation}
\begin{aligned}
\pm\varepsilon(\vb{k}) &= \pm \sqrt{h_1^2 + h_2^2}  - \varepsilon_F\\
&= \pm \sqrt{h_A h_B\delta^{AB}} - \varepsilon_F\\
&= \pm \hbar v_F \sqrt{k_\mu k_\nu e_A^{\mu} e_B^{\nu}\delta^{AB}} - \varepsilon_F\\
&= \pm \hbar v_F \sqrt{k_\mu k_\nu g^{\mu\nu}} - \varepsilon_F,
\end{aligned}
\end{equation}
where we have introduced the effective metric tensor $g^{\mu\nu} = e_A^{\mu} e_B^{\nu} \delta^{AB}$ that captures the anisotropic deformation of the Fermi surface induced by the frame field.
At the Fermi energy, the electron states form a Fermi surface defined by the condition $\varepsilon(\vb{k}) = 0$, which can be compactly written as
\begin{equation}
g^{\mu\nu}k_\mu k_\nu = \frac{\varepsilon_F^2}{\hbar^2 v_F^2}.
\end{equation}
Thus, the frame field modifies the Fermi surface through the induced metric tensor.

Explicitly, the metric tensor resulting from the phonon-induced frame field takes the form:
\begin{equation}
g^{\mu\nu} = \begin{bmatrix}
(1-\frac{\beta u}{a})^2\cos^2\theta + \sin^2\theta & (1-\frac{\beta u}{a})^2\cos\theta\sin\theta - \sin\theta\cos\theta\\
(1-\frac{\beta u}{a})^2\sin\theta\cos\theta - \cos\theta\sin\theta & (1-\frac{\beta u}{a})^2\sin^2\theta + \cos^2\theta
\end{bmatrix}.
\end{equation}
In the absence of phonon displacement ($u = 0$), the metric reduces to the identity matrix:
\begin{equation}
g^{\mu\nu, \text{equi}} = \begin{bmatrix}
1 & 0\\
0 & 1
\end{bmatrix},
\end{equation}
which corresponds to an isotropic, circular Fermi surface.
Treating the phonon displacement $u$ as a small perturbation, we expand the metric tensor to linear order in $u$:
\begin{equation}
\delta g^{\mu\nu} = -\frac{2\beta u}{a}\begin{bmatrix}
\cos^2\theta & \cos\theta\sin\theta\\
\sin\theta\cos\theta & \sin^2\theta
\end{bmatrix}.
\end{equation}
The deformation of the Fermi surface encoded in $\delta g^{\mu\nu}$ can be decomposed into two distinct components:
\begin{equation}
\delta g^{\mu\nu} = \delta g^{\mu\nu,(0)} + \delta g^{\mu\nu,(2)},
\end{equation}
where $\delta g^{\mu\nu,(0)}$ is the isotropic part corresponding to a uniform dilation or contraction of the Fermi surface (angular momentum $l = 0$):
\begin{equation}
\delta {g}^{\mu\nu, (0)} = -\frac{\beta u}{a}\begin{bmatrix}
1 & 0\\
0 & 1
\end{bmatrix},
\end{equation}
and $\delta g^{\mu\nu,(2)}$ is the traceless part corresponding to an anisotropic, elliptic deformation with angular momentum $l = 2$:
\begin{equation}
\delta {g}^{\mu\nu, (2)} = -\frac{\beta u}{a}\begin{bmatrix}
\cos(2\theta) & \sin(2\theta)\\
\sin(2\theta) & -\cos(2\theta)
\end{bmatrix}.
\end{equation}

This quadrupolar component, $\delta g^{\mu\nu,(2)}$, describes a nematic distortion of the Fermi surface that preserves its area but breaks rotational symmetry, transforming a circular Fermi surface into an elliptical one.
Such a deformation is essential for the phonon to couple to the Hall viscosity channel, as it generates the necessary $l = 2$ harmonic of the Fermi surface that gives rise to the nondissipative transverse viscous stress.

\section{Phonon dynamics}

In this section, we show how the phonon dynamics is modified by the Hall viscosity of the electron fluid when the phonon acts as a dynamic frame field.
In the long-wavelength limit, the phonon displacement field $\vb{u}$ can be described by the following effective action for an optical phonon mode
\begin{equation}
S_{0}[\vb{u}] = \frac{1}{2}\int dt d^2x\, \rho_I \qty(\dot {\vb{u}}^2 - \omega_0^2\vb{u}^2),
\end{equation}
where $\rho_I$ is the ion mass density, $\omega_0$ is the bare phonon frequency, and $\vb{u}$ is the in-plane phonon displacement field.

When the phonon modulates the local electronic geometry via the frame field, its effect can be encoded in the coframe field $e^A_\mu$, defined as the inverse of the frame field $e^\mu_A$, satisfying $e^A_\mu e^\mu_B = \delta^A_B$.
To leading order in the phonon displacement, the coframe field can be calculated by matrix inversion:
\begin{equation}
\begin{pmatrix}
e_x^1 & e_x^2 \\
e_y^1 & e_y^2
\end{pmatrix}=
\begin{pmatrix}
e_1^x & e_1^y \\
e_2^x & e_2^y
\end{pmatrix}^{-1} \approx \begin{bmatrix}
\qty(1+\frac{\beta u}{a})\cos\theta & -\sin\theta\\
\qty(1+\frac{\beta u}{a})\sin\theta & \cos\theta
\end{bmatrix}.
\end{equation}
Here, the coframe field can be separated into an equilibrium part and a perturbation induced by the phonon,
\begin{equation}
e^A_\mu = \delta^A_\mu + w^A_\mu = \begin{bmatrix}
\cos\theta & -\sin\theta\\
\sin\theta & \cos\theta
\end{bmatrix} + \begin{bmatrix}
\frac{\beta u}{a}\cos\theta & 0\\
\frac{\beta u}{a}\sin\theta & 0
\end{bmatrix}.
\end{equation}
Substituting this into the effective Hall viscosity action for the frame field, we obtain
\begin{equation}
\begin{aligned}
S_{H}[\vb{u}] &= \frac{\eta_H}{2}\int dt d^2x\, \epsilon^{\mu\nu\rho}w_\mu^A \partial_\nu w_\rho^B \delta_{AB}\\
&= \frac{\eta_H\beta^2}{2a^2}\int dt d^2x\, \epsilon^{\mu\nu\rho} u_\mu \partial_\nu u_\rho,
\end{aligned}
\end{equation}
where $\eta_H$ is the electron Hall viscosity coefficient, and $\epsilon^{\mu\nu\rho}$ is the Levi-Civita tensor.
Because $\mu,\rho$ are restricted to spatial components ($x,y$) and $\epsilon^{\mu\nu\rho}$ is fully antisymmetric, only the temporal derivative term survives with $\nu = t$.
Thus, the Hall viscosity term contributes an additional term that couples the phonon velocity and displacement via a time-reversal symmetry breaking term.

The total phonon effective action is then given by
\begin{equation}
S_{\text{eff}}[\vb{u}] = S_{0}[\vb{u}] + S_{H}[\vb{u}].
\end{equation}
From this, the effective Lagrangian becomes
\begin{equation}
\mathcal{L}_{\text{eff}} = \frac{1}{2}\rho_I \qty(\dot {\vb{u}}^2 - \omega_0^2\vb{u}^2) + \frac{\eta_H\beta^2}{2a^2} \vb{u} \times \dot{\vb{u}}.
\end{equation}
The phonon equation of motion is obtained by varying this effective action
\begin{equation}
\rho_I \qty(\ddot{\vb{u}} + \omega_0^2\vb{u}) = \frac{\eta_H\beta^2}{2a^2}\dot{\vb{u}} \times \hat z.
\end{equation}
This equation describes a phonon mode subjected to a transverse force, leading to a lifting of degeneracy between left-handed and right-handed circularly polarized modes.
Solving the equation of motion in the chiral basis defined by $u_{\pm} \equiv (u_x \pm i u_y)/\sqrt{2}$, the phonon frequencies are found to be:
\begin{equation}
\label{phonon frequencies}
\omega_{\pm} = \sqrt{\omega_0^2 + \left(\frac{\beta^2}{2a^2\rho_I}\eta_H\right)^2}\pm\frac{\beta^2}{2a^2\rho_I}\eta_H.
\end{equation}
Thus, the phonon frequency splitting is given by
\begin{equation}
\delta\omega = \frac{\beta^2}{a^2\rho_I}\eta_H,
\end{equation}
which is directly proportional to the electron Hall viscosity.
This result reveals how the phonon dynamics, in the presence of an emergent frame field, provides a direct probe of the Hall viscosity through a measurable splitting of the chiral phonon modes.

\section{Computational details}
\subsection{First-principles methods}
In this section, we present numerical details on the electron band structures, phonon spectra, and electron-phonon coupling. 
All DFT calculations are performed within the \textsc{Quantum Espresso} package~\cite{QUANTUM_ESPRESSO}.
For Cd$_3$As$_2$, we employ the optimized norm-conserving Vanderbilt (ONCV) pseudopotential~\cite{hamann2013optimized} and the Perdew-Burke-Ernzerhof (PBE) functional~\cite{perdew1996generalized}. 
Atomic structural parameters are obtained from Ref.~\cite{The-Crystal}.
An energy cut-off of 100 Ry and a $k$-grid of $2\times2\times2$ are used for self-consistent calculations. The electronic band structure is shown in Fig. S1(a). 

Phonon dispersions and eigenvectors are calculated based on the finite-displacement approach using the \textsc{Phononpy} package~\cite{togo2023first,togo2023implementation}. 
We construct two linearly polarized modes from the doubly degenerate $E_u$ modes at the $\Gamma$ point and displace atoms along the phonon eigenvectors with an amplitude of 1 \AA. 
Due to the considerable computational cost, we initially use a $k$-grid of $11\times11$ in DFT calculations to obtain the in-plane contour plot of the Dirac cone perpendicular to the $\Gamma$-$Z$ direction as in Fig. S1 (b) and (c), subsequently interpolating the map to a $100\times100$ $k$-grid.

\begin{figure}
\setcounter{figure}{0}
\renewcommand{\figurename}{Fig.}
\renewcommand{\thefigure}{S\arabic{figure}}
    \centering
    \includegraphics[width=\linewidth]{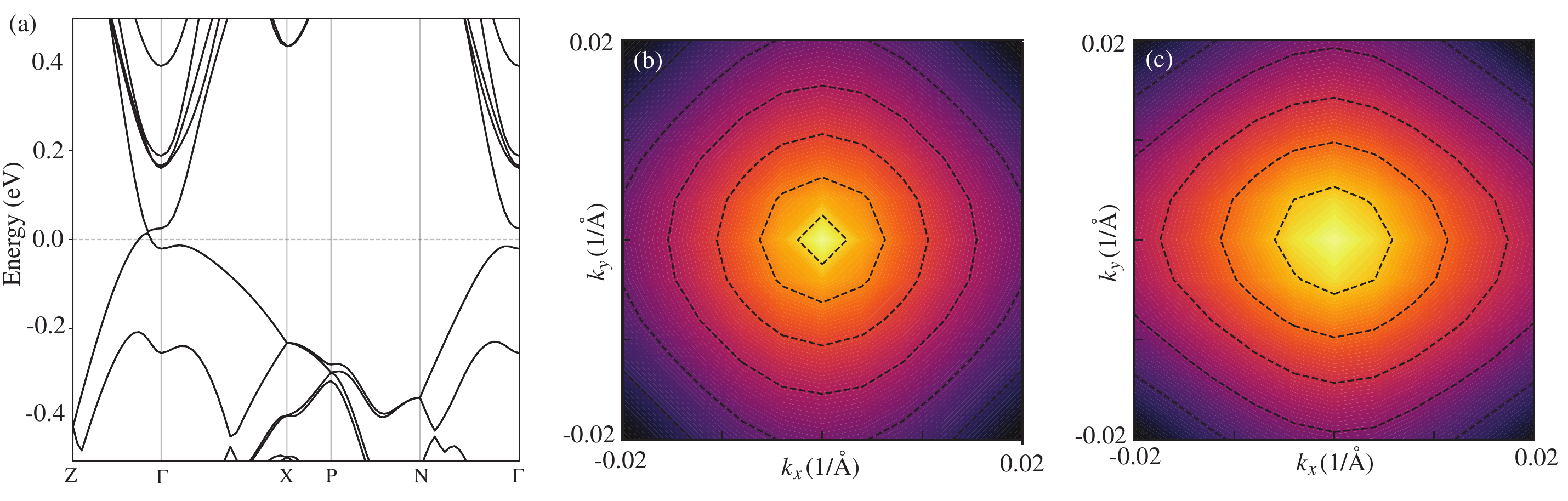}
    \caption{(a)Electronic band structure of $\text{Cd}_3\text{As}_2$ showing one Dirac point $\boldsymbol{K}_+$ on the $\Gamma$-Z axis. (b)(c) In-plane contour plots of the Dirac cone perpendicular to the $\Gamma$-$Z$ direction on a $k$-grid of $11\times11$ calculations using the DFT in equilibrium and in the presence of the $\Gamma_{8}$($E_u$) mode. }
    \label{fig:enter-label}
\end{figure}

\subsection{Electron-phonon coupling in Cd$_3$As$_2$}
We start from the many-body Hamiltonian of electrons and ions at equilibrium positions, $H = H_{\text{e}} + H_{\text{ion}} + H_{\text{ei}}$, where $H_{\text{ei}} = \sum_{i\vb{R}\alpha} V_{\text{ei}} (\vb{r}_i-\vb{R}-\bm{\tau}_\alpha)$ is the electron-ion interaction Hamiltonian. 
Here $\vb{R}$ is the lattice vector for unit cells and $\bm{\tau}_\alpha$ is the position of ion $\alpha$ within a unit cell. 
If the ions vibrate, $H_{\text{ei}}$ will change to $\sum_{i\vb{R}\alpha} V_{\text{ei}} (\vb{r}_i-\vb{R}-\vb{\tau}_\alpha + \vb{u}_{\vb{R}\alpha})$, where $\vb{u}_{\boldsymbol{R}\alpha}$ is the atomic displacement of the ion $\alpha$ in the unit cell $\vb{R}$ from its equilibrium position. 
It is usually small compared to the lattice constant, allowing us to expand the potential to the first order of $\vb{u}_{\vb{R}\alpha}$ as $\sum_{i\vb{R}\alpha} V_{\text{ei}}(\vb{r}_i - \vb{R} - \bm{\tau}_\alpha) + \vb{u}_{\vb{R}\alpha} \cdot \bm{\nabla} V_{\text{ei}}(\vb{r}_i - \vb{R} - \boldsymbol{\tau}_\alpha)+ O(\vb{u}_{\boldsymbol{R}\alpha}^2)$. Thus the electron-phonon coupling Hamiltonian comes from the first-order correction,
\begin{equation}
    H_{e\text{-ph}} = \sum_{i\vb{R}\alpha} \vb{u}_{\vb{R}\alpha} \cdot \bm{\nabla} V_{\text{ei}}(\vb{r}_i - \vb{R} - \vb{\tau}_\alpha) = \int d\vb{r} \rho(\vb{r}) \sum_{\vb{R}\alpha} \delta\vb{u}_{\vb{R}\alpha} \cdot \bm{\nabla} V_{\text{ei}}(\vb{r}_i - \vb{R} - \bm{\tau}_\alpha),
\end{equation}
where $\rho(\vb{r})$ is the density operator of electrons, and $\rho(\vb{r}) = \sum_{i} \delta(\hat{\vb{r}}_i - \vb{r})$ in the coordinate representation. 
Now we move to the second quantization of density operator $\rho(\vb{r}) = \psi^\dagger(\vb{r})\psi(\vb{r})$. 
The field operator projected to Bloch electronic basis is $\psi^\dagger(\vb{r}) =  \sum_{n\vb{k}}\psi^*_{n\vb{k}}(\vb{r}) c^\dagger_{n\vb{k}}$, where $n$ is the energy band index and $\vb{k}$ is electronic wavevector. 
The Bloch wavefunctions are given by $\psi_{n\vb{k}}(\vb{r}) = u_{n\vb{k}}(\vb{r})e^{i\vb{k}\cdot\vb{r}}$, where $u_{n\vb{k}} = \braket{\vb{r}}{n\vb{k}}$ is the periodic Bloch function within one unit cell. 
Under this representation, the electron-phonon coupling Hamiltonian becomes
\begin{equation}
    H_{e\text{-ph}} = \sum_{nm} \sum_{\vb{k}\vb{k}'} \int d\vb{r} \psi^*_{n\vb{k}}(\vb{r}) \sum_{\vb{R}\alpha} \vb{u}_{\vb{R}\alpha} \cdot \bm{\nabla} V_{\text{ei}}(\vb{r}_i - \vb{R} - \bm{\tau}_\alpha) \psi_{m\vb{k}'}(\vb{r})  c^\dagger_{n\vb{k}}c_{m\vb{k}'}.
\end{equation}
Now we expand the atomic displacement of the ion at $\vb{R}+\bm{\tau}_\alpha$ in terms of the vibrational modes,
\begin{equation}
    \label{disp_expsn}
    \vb{u}_{\vb{R}\alpha} = \sum_{\vb{q},\nu} \sqrt{\frac{M_0}{NM_{\alpha}}}e^{i\vb{q} \cdot \vb{R}}\ \bm{\xi}_{\alpha}^{(\nu)}(\vb{q}) u_{\vb{q}}^{(\nu)},
\end{equation}
where $\vb{q}$ is the phonon wavevector, $\nu$ labels the normal modes, $\omega_{\vb{q}}^{(\nu)}$ is the frequency dispersion, and $M_0$ is an arbitary reference mass (typicall chosen to be the proton mass), and $u_{\boldsymbol{q}}^{(\nu)}$ is the complex normal coordinate of the displacement projected onto the normal mode $\nu$. 
The column vector $\vb{\xi}_{\alpha}^{(\nu)}(\vb{q})$ is the normal mode eigenvector that solves the dynamic matrix problem, $\sum_{\alpha'} D_{\alpha\alpha'}(\vb{q}) \bm{\xi}_{\alpha'}^{(\nu)}(\vb{q}) = \omega_{\vb{q}}^{(\nu)2} \bm{\xi}_{\alpha}^{(\nu)}(\vb{q})$, 
The quantization of the phonons is given by 
\begin{equation}
    u_{\vb{q}}^{(\nu)} = l^{(\nu)}_{\vb{q}} [a^{(\nu)}_{\vb{q}} + a^{(\nu)\dagger}_{-\vb{q}}] = \sqrt{\frac{\hbar}{2M_0\omega^{(\nu)}_{\vb{q}}}} [a^{(\nu)}_{\vb{q}} + a^{(\nu)\dagger}_{-\vb{q}}],
\end{equation}
where $l^{(\nu)}_{\vb{q}}$ is the "zero-point" displacement amplitude. We arrives at the Frölich Hamiltonian for the electron-phonon coupling,
\begin{equation}
\begin{aligned}
  H_{e\text{-ph}} = \frac{1}{\sqrt{N}}\sum_{nm\vb{k}} \sum_{\vb{q},\nu} g_{mn}^{(\nu)}(\vb{k},\vb{q}) Q_{\vb{q}}^{(\nu)}  c^\dagger_{n\vb{k}+\frac{\vb{q}}{2}}c_{m\vb{k}-\frac{\vb{q}}{2}},
\end{aligned}
\end{equation}
where the electron-phonon coupling matrix element is given by
\begin{equation}
    g_{mn}^{(\nu)}(\vb{k},\vb{q}) =  \bra{n\vb{k}+\frac{\vb{q}}{2}} ({\nabla}V_{\text{ei}})_{\vb{q}}^{(\nu)} \ket{m\vb{k}-\frac{\vb{q}}{2}}_{\text{uc}}.
\end{equation}
We have defined $({\nabla}V_{\text{ei}})_{\vb{q}}^{(\nu)} = \sum_{\vb{R}\alpha}\sqrt{M_0/M_{\alpha}} e^{-i\vb{q} \cdot (\vb{r} - \vb{R})} \bm{\xi}_{\alpha}^{(\nu)}(\vb{q}) \cdot \bm{\nabla}V_{\text{ei}}(\vb{r} - \vb{R}-\bm{\tau}_\alpha)$.

Since we are interested in the low-energy effective $e$-ph coupling in the form of Eq. (2) in the main text, we adopt the finite-displacement approach in the first-principles calculations.
We make the following approximation for phonons in Cd$_3$As$_2$: (i) We consider the long-wavelength limit ($\vb{q} = 0$), where the atomic displacements are
\begin{equation}
    \vb{u}_{\alpha} = \sqrt{\frac{M_0}{M_{\alpha}}} \bm{\xi}_{\alpha}^{(\nu)} l^{(\nu)} u_0^{(\nu)},
\end{equation}
where only the zero-center contribution is kept and we only consider one phonon mode $\nu$.
(ii) Since the tight-binding model used for Cd$_3$As$_2$ has only one $s$ and one $p$ orbital per site while the real lattice has 80 atoms per unit cell, we take the average of $|\vb{u}_{\alpha}|$ over the index $\alpha$:
\begin{equation}
    |\vb{u}| = |\frac{1}{N_{\alpha}}\sum_{\alpha}\vb{u}_{\alpha}| = |\frac{1}{N_{\alpha}}\sum_{\alpha} \sqrt{\frac{M_0}{M_{\alpha}}} \bm{\xi}^{(\nu)}_\alpha l^{(\nu)}| \approx |\frac{1}{N_{\alpha}}\sqrt{\frac{M_0}{\overline{M}}}l^{(\nu)}\sum_{\alpha} \bm{\xi}^{(\nu)}_\alpha|
\end{equation}
where $\overline{M}$ is the average mass of the ions in the unit cell.
For the $E_u$ mode we calculated, the amplitude $\sqrt{\frac{M_0}{\overline{M}}}l^{(\nu)}$ is set to be $1$ \AA. 
The weighted average of the displacement magnitude is $|\vb{u}| \approx 0.0019$ \AA. 
Then we can estimate the magnitude of the frame field $e^A_\mu$ from the deformed Dirac cone. 
The ratio of the Dirac cone slope between $x$ and $y$ axes is about $\beta |\vb{u}|/a \approx 1.211$.
Thus we can make an estimate of the electron-phonon coupling parameter $\beta/a \approx 632$ \AA$^{-1}$.

\end{document}